\documentclass[pageno]{jpaper}

\usepackage[normalem]{ulem}
\usepackage{tabularx, booktabs, multirow, amsmath}

\begin{document}

\title{
Evaluating Ternary Adders using a hybrid Memristor / CMOS approach }

\date{}

\author{Dietmar Fey\\
Friedrich-Alexander-University Erlangen-N{\"u}rnberg (FAU) \\
Chair Computer Architecture \\
Martensstr. 3, 91058 Erlangen, Germany \\
dietmar.fey@fau.de}

\maketitle

\thispagestyle{empty}

\begin{abstract}
\vspace{1.2mm}
This paper investigates the potentials of using a hybrid memristor CMOS technology, called MeMOS, for the realisation of ternary adders. Ternary adders exploit the qualitative advantage of multi-value storage capability of memristors compared to conventional CMOS flip-flops storing only binary values in one cell. Furthermore they carry out an addition in $O(1)$ and are therefore considered. The MeMOS approach is compared to a CMOS solution for the ternary adders using multi value memristors as registers concerning the achievable latency and the energy consumption. It is shown that using the TEAM, VTEAM model and a model considering commercially available memristors from Known the approach of using CMOS ternary adders using memristors as multi-value register memory is to prefer. MeMOS circuits have advantages for a static operation mode, i.e. if they are operated after a reset. 
\end{abstract}

\section{Introduction}

%This document provides the formatting instructions for submissions to the 25th International Conference on Parallel Architectures and Compilation Techniques, 2016. In an effort to respect the efforts of reviewers and in the interest of fairness to all prospective authors, we request that all submissions to PACT'16 follow the formatting and submission rules detailed below.  Submissions that violate these instructions may not be reviewed, at the discretion of the program chair, in order to maintain a review process that is fair to all potential authors.  This is a generous format, with plenty of space -- there should be no need to tweak it in any significant way.

%An example submission (formatted using the PACT'16 submission format) that contains the submission and formatting guidelines can be downloaded from the conference submission website at
%{\em http://pactconf.org/files/2016/02/pact16format.tar.gz}.

%All questions regarding paper formatting and submission should be directed to the program chairs.

New memory technologies, like e.g. memristors, offer the possibility of so-called in-memory computing concepts. Characteristic for such concepts is a paradigm change in the way how data processing is done. Whereas in a pure von-Neumann computer data is brought to the instructions in a computing architecture that uses in-memory computing the instructions are brought to the data. Concrete does that mean the memory storage element is not purely used for storage, moreover the storage is an integral part of the processing. 

Using memristors as storage elements in an in-memory computing architecture means that logical resistor networks are built up and the resistor values can be changed to new ones according to the result of a processing step in which the old resistor value was included.         

One of the first concepts in this sense in the world of memristive computing  was the so-called IMPLY logic \cite{Borghetti2010}, \cite{Kvatinsky2011}. Two parallel connected memristors got as input voltage levels that are either shortly before a switching level of the memristor or are above that limit. In dependence of the content of the memristor pair, which is the input of the function, a certain resistor network is built up. Depending on the adjusted resistances in that resistor network the state of one memristor, operating as the output, can be changed or not if a voltage above the threshold limit is applied. By that a kind of operation mode a quasi inherent implementation of the Boolean implication operator is realised.     

First made proposals with that IMPLY logic include basic Boolean functions like NOT, NAND or NOR. However, the disadvantage is that the realisation of such elementary Boolean functions are mapped onto several subsequent executions of IMPLY operations what causes a higher latency. This problem does not occur in a pure CMOS logic in which operations like NAND or NOR are quasi inherent to that processing which is based on complementary switching of PMOS and NMOS transistors. This means that the principal advantage of in-memory computing concepts over CMOS like less energy consumption and possibly less latency, which is given due to the fact that storing and processing takes place in the same device, is lost.    

New expanded proposals like so-called MAGIC \cite{MAGICGates} and MAD gates \cite{MADGates} or also the concepts presented in \cite{Gao2013} and of MeMOS \cite{DBLP:journals/corr/Singh15b}, the last one stands for combining of memristors and CMOS,   overcome the mentioned disadvantage and allow to carry out basic Boolean operators in one elementary step. This paper is focused on using the MeMOS concept since the requirements to the control path for the memristors seems to be less complicate and therefore one can expect even a short- or mid-term solution concerning the realisation of real circuits compared to the other solutions. The control path in MeMOS is more or less reduced to CMOS inverter gates what is much easier compared to the more difficult driver circuitry that MAD gates require. Of course, the price one has to pay for the easier realisation of MeMOS is a higher number of memristors as needed for MAD gates and MAGIC.  

All of the three concepts mentioned above allow to do the next necessary step in digital memristive computing for the community, namely to move away from simple Boolean gates towards more sophisticated memristive arithmetic circuits. According to the conviction of the authors this step is absolutely necessary if memristors will have a chance to be used as digital logic any time in the future in order to exploit its principal benefits like less energy consumption and may be also less latency. 

In addition, we would like to focus that the success of digital memristive computing requires also to exploit features offered by new memory technologies like memristors which are not possible with pure CMOS storing elements. This concerns e.g. the multi-value storing capabilities of some NVMs in general, and in particular of memristors. We favour in this presented work a ternary number representations realised in memristors as another focus aside the MeMOS concept. 

In this sense we would like to investigate if memristor technology has the potential to realise ternary computers in the future. The reason why we favour ternary arithmetic than binary is the advantage of ternary structures to carry out an addition in a constant number of steps independent of the operand's word length , i.e. in $O(1)$. On the contrary a binary adder works best at $log(N)$ by an area increase that is limited to $(N \, log(N)$, if $N$ is the number of used bits.  

In order to prove the potential and to quantify what has to be done in the future to bring digital memristive computing to success we made a comparative investigation concerning required processing time and energy consumption for ternary adders realised in three different technologies.  (i) A pure CMOS solution of the adders using two flip-flops to store one trit, i.e. a ternary digit; (ii) A non in-memory computing approach using memristors, i.e. memristor are just used as multi-value memory in a digital CMOS circuit; (iii) A MeMOS solution for those adders.  

In order to obtain the energy and latency values for memristor based adder architectures, i.e. (i) and (ii), we used an own simulation environment written in C++ that solves the memristor behaviour by a discrete solution of the differential equations for different models proposed in literature, namely the TEAM \cite{Kvatinsky_TEAM} and the VTEAM \cite{Kvatinsky_VTEAM} model. The difference between TEAM and VTEAM model is that the last one considers explicitly a threshold at which the I/U curve for a memristor starts to show a slope. Furthermore, we consider a statistical model for one of the first commercially available memristors, namely the memristors from Knowm \cite{Knowm}. 

The remainder of the paper is organized as follows. Chapter 2 presents an introduction in ternary arithmetic and explains the specific adders we selected for this investigation. Next chapter briefly introduces the MeMOS concept and shows the MeMOS solution for our preferred ternary adder. Chapter 4 introduces the simulation system we have developed for the evaluation of MeMOS circuits. Chapter 5 continues with an evaluation for the ternary adders in MeMOS and discusses the pro and cons. Finally we end the paper with a conclusion.     

\section{Ternary adders}

We have investigated two adders operating on ternary numbers which are processed by binary logic for a realisation in MeMOS technology. The two ternary adders are operating both to a base $r=2$. One adder, denoted in the following as $base\_2\_step\_3$ adder, requires three discrete time steps for the processing, but the price is that this adder shows a more complex logic per steps compared to the other adder, denoted as $base\_2\_step\_4$, which requires however 4 discrete time steps. These two adders are more promising than other adders offering also non-binary operand presentation and constant addition time. E.g. there exist adders, that show only two discrete steps but require a much more complex logic what is caused by the fact that these adders are using operands to a base $r \geq 2$. Details concerning this computer arithmetic stuff can be found in \cite{GonzalezM00}, \cite{MemSysPaper}.

Both adders are using for their operands a so-called signed digit (SD) number representation, i.e. they allow not only positive but also negative digits. Concretely, a so-called balanced ternary number representation to base $r=2$ comes into use. That means, generally for each digit $sd_i$ of a SD number $sd$ holds $sd_i \in \lbrace-r+1, ..., r-1\rbrace$, and for the case $r = 2$, $sd_i \in \lbrace-1,0,1\rbrace$. The value of $sd$, having a digit length $N$, is then calculated according to (\ref{eq:value_sd}). 

\begin{equation}
\label{eq:value_sd}
  sd =  \sum_{i=0}^{N} sd_i \cdot r^i  
\end{equation} 

Due to the signed digits we receive redundant representations, i.e. two different SD numbers possess the same value, e.g. $1 0 \overline{1} 1 = (7)_{10} = 1 0 0 \overline{1}, \overline{1} \equiv -1$. 

As next we explain briefly the working principle of the two adders. We will start with the adder $base\_2\_step\_4$ which carries out an addition in four steps independent of the word length. 

For the binary processing of the ternary SD number we have to use a coding for each trit. The coding we are using is a so-called (negative, positive) coding, abbreviated as (n,p)-coding, shown in Table \ref{table:trit_coding}. The (n,p)-coding has the advantage that the inverse $-sd_i$ of a digit $sd$ can be easily formed by exchanging the negative and the positive part of $sd$. 

\begin{table}[h!]
  \centering
  \begin{tabular}{|p{3cm}|p{3cm}|}
    \hline
    \textbf{Ternary SD digit} \newline (trit)  & \textbf{digital 2-bit coding} \newline (neg  pos)\\
    \hline
    \hline
    -1 &   1  0\\
    \hline
    0  &   0  0\\ 
    \hline
    1  &   0  1\\ 
    \hline
  \end{tabular}
  \caption{Binary coding of the trits used for the digital processing. The first bit is interpreted as positive, the second one as negative.}
   \label{table:trit_coding}
\end{table}

The $base\_2\_step\_4$ adder consists of two subsequent connected subadders, which expect one operand as binary input and the other one as an SD number. This makes the gate logic more easier. An example for an addition that is carried out with such a subadder is shown in Table  \ref{table:Add_SD_Binary}. 

\begin{table}[htbp]
\centering%
\begin{tabular}{rccccll}
x \  = & (1 & -1 & 1 & 1)$_2$ & $= \ (7)_{10}$ \\
+y \ = & (0 &  1 & 1 & 0)$_2$ & $= \ (6)_{10}$ \\ 
\hline
\vspace{6pt}
step 1:\\
      & -1 & 0 & 0 & -1 & $= \ z$ \\
         1 &  0 & 1 & 1 &  0 & $= \ c \ \text{shifted to left}$ \\
\hline
\vspace{6pt}
step 2:\\
     1 & -1 & 1 & 1 & -1 & = $s = 1 \cdot 2^4 - 1 \cdot 2^3 + 1 \cdot 2^2$\\
       & & & & & $  + \ 1 \cdot 2^1 -1 \cdot 2^0 = 13$ \\ 
\hline
\end{tabular}
\caption{Addition of $SD$ number $x$ with a binary number $y$.}
\label{table:Add_SD_Binary}
\end{table} 

Assuming the (n,p) coding we will get the following Boolean equations, (\ref{eq:first_c}) to (\ref{eq:sum_minus}), for the $base\_2\_step\_4$ adder that have to be applied to calculate the intermediate binary values $c_i^+$, $z_i^-$, $s_i^+$, and $s_i^-$ in each digit position of the adder. 

\begin{align}
\label{eq:first_c}
c_i^+&=x_i^+ \vee \left( y_i \wedge \overline{x_i^-} \right) \\ 
\label{eq:first_z}
z_i^-&=\left( x_i^+ \vee x_i^- \right) \oplus y_i  \\
\label{eq:sum_plus}
s_i^+&=\overline{z_i^-} \wedge c_{i-1}^+\\ 
\label{eq:sum_minus}
s_i^-&=\overline{c_{i-1}^+} \wedge z_i^-  
\end{align}

\begin{figure}[htbp]
\centering
\includegraphics[width=0.98\columnwidth]{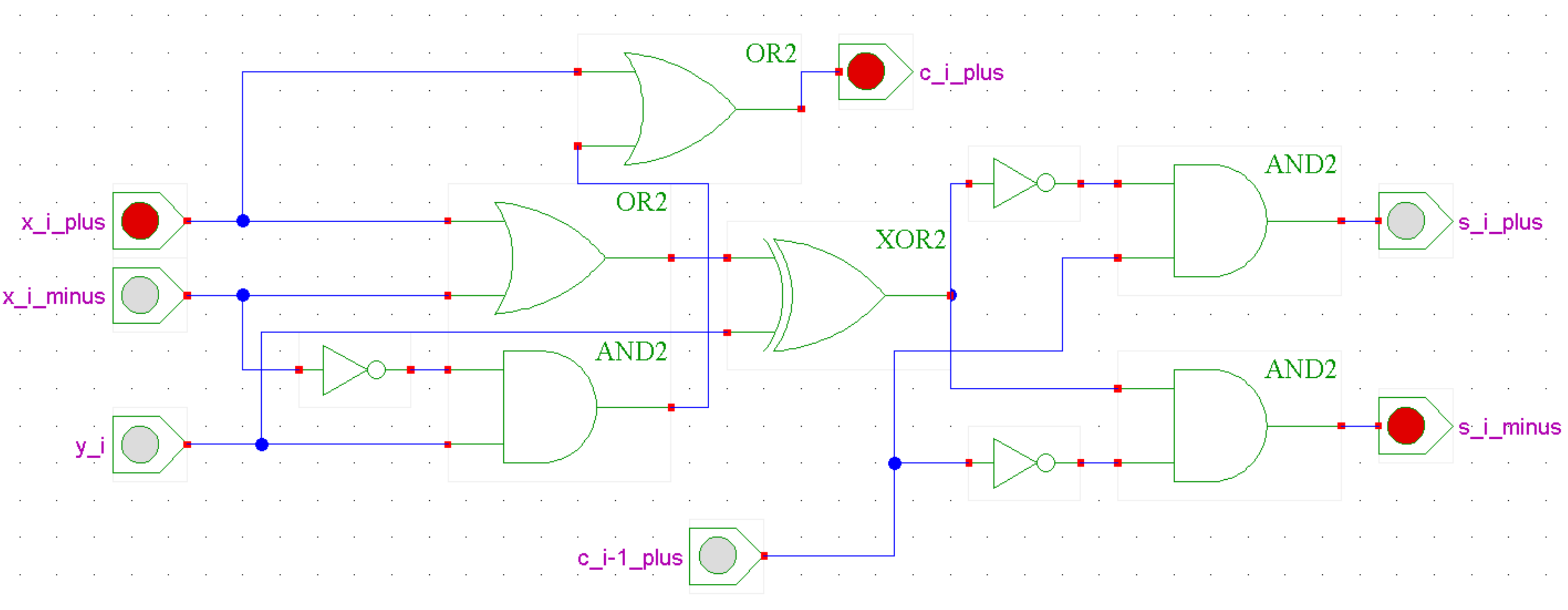}
\caption{Gate layout for the processing of one digit in the ternary $Radix\_2\_step\_4$ adder.}
\label{fig:base_2_step_4_adder_cell}
\end{figure}

Fig. \ref{fig:base_2_step_4_adder_cell} shows a gate layout for one basic building block for this adder for the processing of one digit. It is operated on one digit of SD operand $sd1_i$, i.e. on its positive and negative part denoted as $x\_i\_plus$, resp. $x\_i\_minus$, and on the positive part of a digit of the other SD operand $sd2_i$, denoted as $y\_i$.  

In order to process two SD numbers as inputs the addition process has to be be split in two phases (\ref{eq:SD_SD_Base2_Addition}), due to the fact that one subadder of the the $base\_2\_step\_4$ adder has to be fed with one SD input and one binary input. Therefore, first an addition of the first operand $sd_1$ takes place with the positive part $sd^+_2$ of the second input operand $sd_2$. Then, a subsequent subtraction with the negative part $sd^-_2$ has to follow. Both input operands for the subadders are binary. 

\begin{equation}
\label{eq:SD_SD_Base2_Addition}
  sd_1 + sd_2 = (sd_1 + sd_2^+)-sd_2^- \\
\end{equation}

Eq. (\ref{eq:subtract}) shows how a subtraction of an SD number, $sd$, and a binary, $B$, can be easily reduced to an addition. The positive and the negative part of $sd$ have to be exchanged, then the addition with $B$ can be carried out. The result is an SD number, whose positive and negative part  have to be exchanged to form the inverse. 

\begin{equation}
\label{eq:subtract}
sd-B=\left( -1 \right) \cdot \left( \left( -1 \right) \cdot sd + B \right)
\end{equation}

\begin{figure}[htbp]
\centering
\includegraphics[width=0.98\columnwidth]{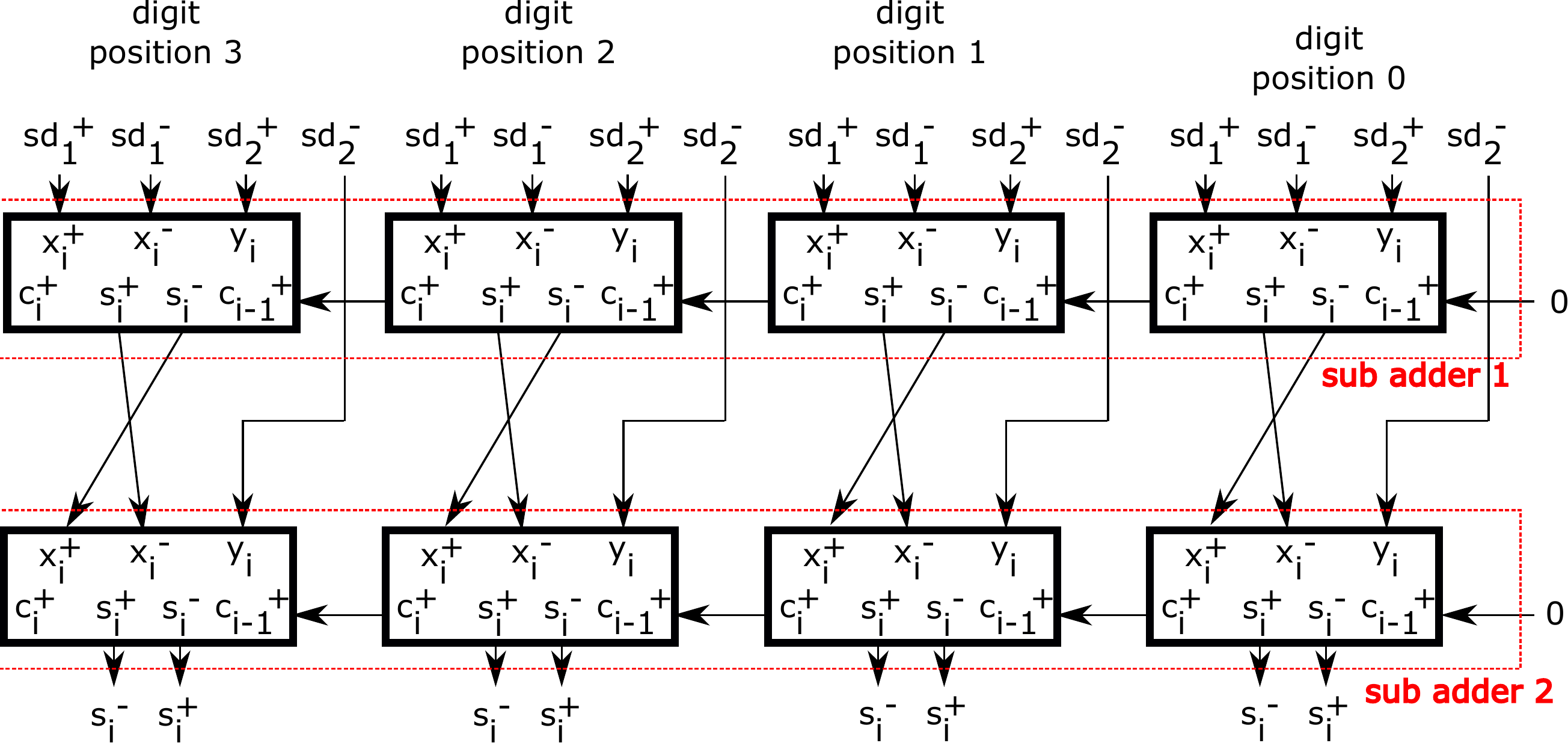}
\caption{Gate layout for the processing of one digit in the ternary $Radix\_2\_step\_4$ adder.}
\label{fig:base_2_step_4_adder}
\end{figure}

Fig. \ref{fig:base_2_step_4_adder} shows a schematic on block level for the complete addition of two $SD$ operands with a width of four digits. Since two subadders are needed we require four steps in total. 

The $base\_2\_step\_3$ adder requires only three steps, but needs a more complicate Boolean logic what is to explain since both input operands for this adder are SD numbers, i.e. a vector of four bit width is read in. Table \ref{table:Add_SD_SD_3steps} shows an example how the addition takes place in three steps. The corresponding Boolean equations are given in (\ref{eq:ADD_SD_SD_3S_1}) to (\ref{eq:ADD_SD_SD_3S_3}). 

\begin{table}[t]
\centering%
\begin{tabular}{rcccclll}
%\hline
\vspace{6pt}
\\
x \  = & (0 & 0 & -1 & 0)$_2$ & $= \ (-2)_{10}$ \\
+y \ = & (0 & 1 & -1 & 0)$_2$ & $= \ (+2)_{10}$ \\ 
\hline
\vspace{6pt}
step 1:\\
       & 0 & -1 & -2 & 0 & $= \ z$ \\
         0 & 1 &  0 &  0 & 0 & $= \ t \ \text{shifted to left}$ \\
\hline
\vspace{6pt}
step 2:\\ 
      0 &  1 &  1 & 0 & 0 & $= \ z'$ \\
      0 & -1 & -1 & 0 & 0 & $= \ t' \ \text{shifted to left}$ \\        
\hline
\vspace{6pt}
step 3:\\
     0 & 0 & 0 & 0 & 0 & = $s = 0$ \\
\hline
\end{tabular}
\caption{Addition of two $SD$ numbers, $x$ and $y$ to base $r=2$, in three steps using mixed redundant number representations, i.e. in first step -2, -1, and 0 are used, input and output digits are either -1,0, or 1.}
\label{table:Add_SD_SD_3steps}
\end{table}

\begin{align}
\label{eq:ADD_SD_SD_3S_1}
t_{i+1} &= {sd1^+_i} \ \overline {sd2^-_i} \vee \overline {sd1^-_i} \ {sd2^+_i} \\
z^+_i &= sd1_i^- \wedge sd2_i^- \nonumber \\
z^-_i &= \overline {sd1^+_i} \ \overline {sd1^-_i} \ {sd2^-_i} \vee sd1^-_i \ \overline {sd2^+_i} \ \overline {sd2^-_i} \vee \nonumber \\
& \ \ \ \  sd1^+_i \ \overline {sd2^+_i} \ \overline {sd2^-_i} \vee \overline {sd1^+_i} \ \overline {sd1^-_i} \ sd2^+_i \nonumber
\end{align}

\begin{align}
\label{eq:ADD_SD_SD_3S_2}
t'_{i+1} &= (\overline {t_i} \wedge z^-_i) \vee z^+_i \\
z'_i &= (\overline t_i \wedge z_i^-) \vee (t_i \wedge \overline z_i^-) \nonumber 
\end{align}
\begin{align}
\label{eq:ADD_SD_SD_3S_3}
s^+_i &= \overline {t'_i} \wedge z'_i \\
s^-_i &= t'_i \wedge z'_i \nonumber 
\end{align}

\begin{figure}[htbp]
\centering
\includegraphics[width=0.98\columnwidth]{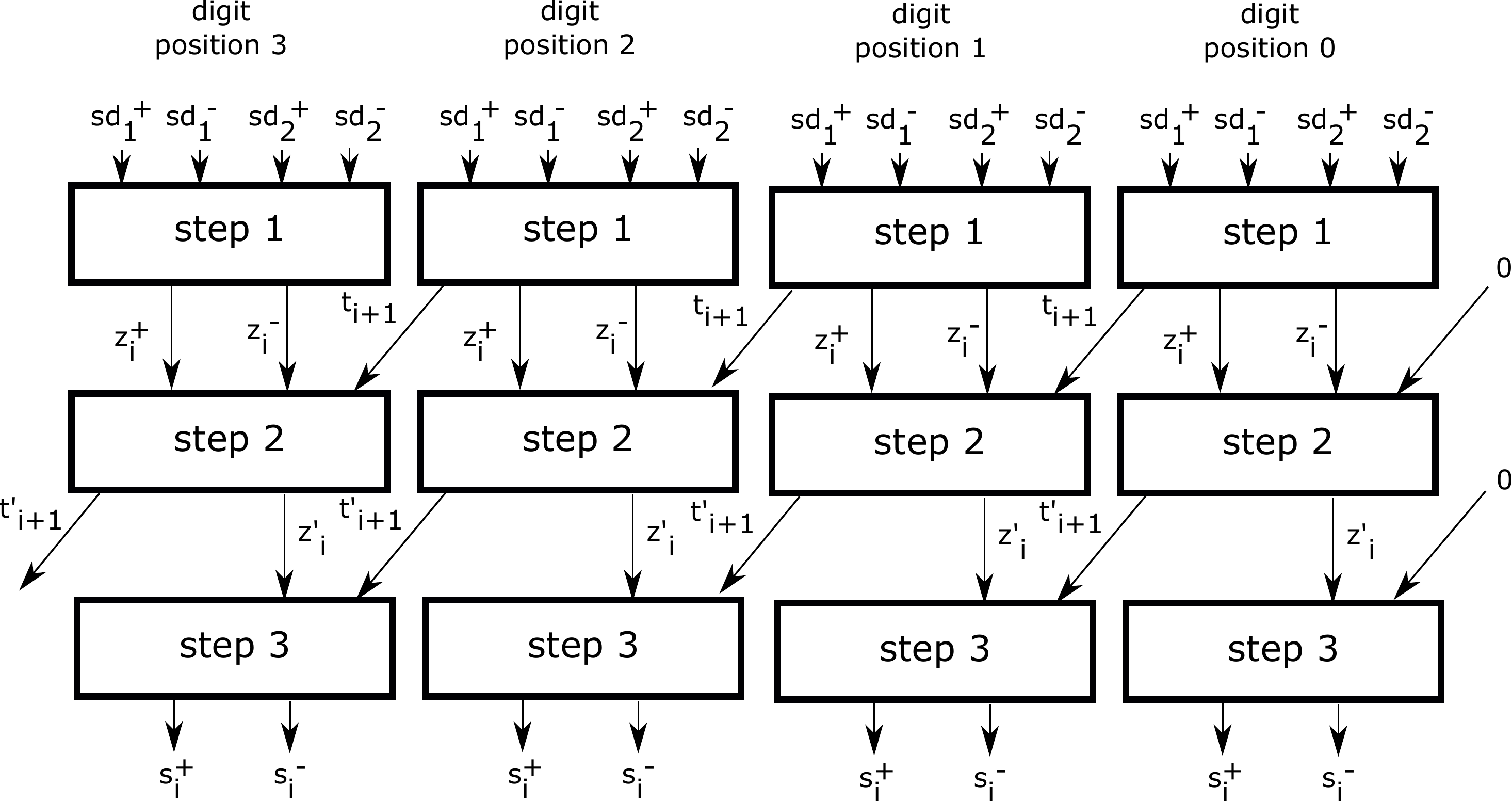}
\caption{Block layout for the processing of two ternary input operands $sd1$ and $sd2$ with the $Radix\_2\_step\_3$ adder.}
\label{fig:base_2_step_3_adder}
\end{figure}

Fig. \ref{fig:base_2_step_3_adder} shows a schematic on block level how the complete adder works. In the blocks denoted \textit{step1}, \textit{step2}, and \textit{step3} the equations (\ref{eq:ADD_SD_SD_3S_1}), (\ref{eq:ADD_SD_SD_3S_2}), and (\ref{eq:ADD_SD_SD_3S_3}) are carried out.

The idea behind the $base\_2\_step\_3$ adder to avoid a carry chain is the following one. In the first step the digits, $z_i$, of the intermediate value $z$, and the transfer bits, $t_i$, of the so-called transfer vector, $t$, are produced. It holds for $z_i \in \lbrace -2,-1,0 \rbrace$, and for $t_i \in  \lbrace 1,0 \rbrace$. Consequently, it is impossible that in the second step a digit with a value equal to $+/-r$ can be produced, i.e. no carry can occur. It requires a final third step to consider the final necessary shift and to produce again an SD number.

The idea behind the $base\_2_step\_3$ adder to avoid a carry is much simpler to understand. The vector $c_i$ contains only 1's and 0's, the vector $z$ contains only $\overline{1}$'s and 0's. Therefor,e it is also impossible that in the second step two $\overline{1}$'s or to 0's meet each other at a certain digit position and no carry can arise.  

Both presented adders have been intensively investigated on technological side by a layout synthesis evaluation by other authors \cite{MemSysPaper} concerning their latency and energy consumption behaviour in comparison to a binary carry-look-ahead adder (CLA). The both adders were connected to a multi-value memristor based register file. As expected for both ternary adders a constant run time was received, about 3\,ns for the $base\_2\_step\_3$ adder and about 3.5\,ns for the $base\_2\_step\_4$ adder for a 130 nm CMOS process from LFoundry.

As expected, too, the run time for a carry-look-ahead (CLA) adder was higher. It started at about 3.6 ns for a bit length of 16 and ended at about 6\,ns for a word length of 512 bits. However, concerning for the energy-delay product the break-even point where the ternary adders were better than the CLA adder was just at 40 digits. We are interested in this paper to see if the energy-delay product would improve if these adders are realised in MeMOS technology. 

\section{MeMOS implementation of ternary adders}

Singh \cite{DBLP:journals/corr/Singh15b} made a proposal to combine memristors and CMOS in order to solve the problem of signal deterioration of the voltage signal level when using a pure ratioed logic \cite{Kvatinsky2012} with memristors. Basic of ratioed logic are voltage dividers based on memristors, which make it difficult to cascade circuits since voltage drops off at the memristors. The idea is to work with CMOS inverters at certain distances in a digital logical memristor circuit to refresh the voltage signals to the level of the supply voltage. Singh denoted this technique as Memristor-CMOS (MeMOS) logic. Further inverters are used if inverted signals are necessary for memristor inputs, which are used as logic gates. 

Basic building blocks of MeMOS are OR and AND circuits in ratioed logic with closing CMOS inverters for signal restoration what establishes a NOR-/NAND logic.

Singh showed benefits for a full adder realised in MeMOS in comparison with a CMOS full adder concerning energy consumption and latency for an assumed TEAM model for the memristors. We want to apply that approach here to more sophisticated circuits, namely ternary adders to profit from further qualitatively benefits offered by memristor technology.

What we need for a perfect mapping onto MeMOS building blocks are Boolean equations formulated as NAND/NOR expressions. In order to receive such expressions it is necessary to expand the Boolean equations by a double inversion. We demonstrate that exemplarily for the calculation of the $c_i^+$ (\ref{eq:first_c}) signal that is generated in the first step of the $base\_2\_step\_4$ adder (\ref{eq:NAND-logic}).

\begin{align}
\label{eq:NAND-logic}
\overline{\overline{c_i^+}} &= \overline{ \overline{x_i^+ \vee \left( y_i \wedge \overline{x_i^-} \right)}}\\
&= \overline{\overline{x_i^+} \wedge \left( \overline{y_i} \vee x_i^- \right)} \nonumber \\ 
&= \overline{ \overline{x_i^+} \wedge \overline{ \left( y_i \wedge \overline{x_i^-} \right) } } \nonumber
\end{align}

%={overline{overline{x_i^+ \vee \left( y_i \wedge \overline{x_i^-}}}} = \overline{\overline{y_i} \wedge x_i_+} \vee x_i-

\begin{figure}[htbp]
\centering
\includegraphics[width=0.98\columnwidth]{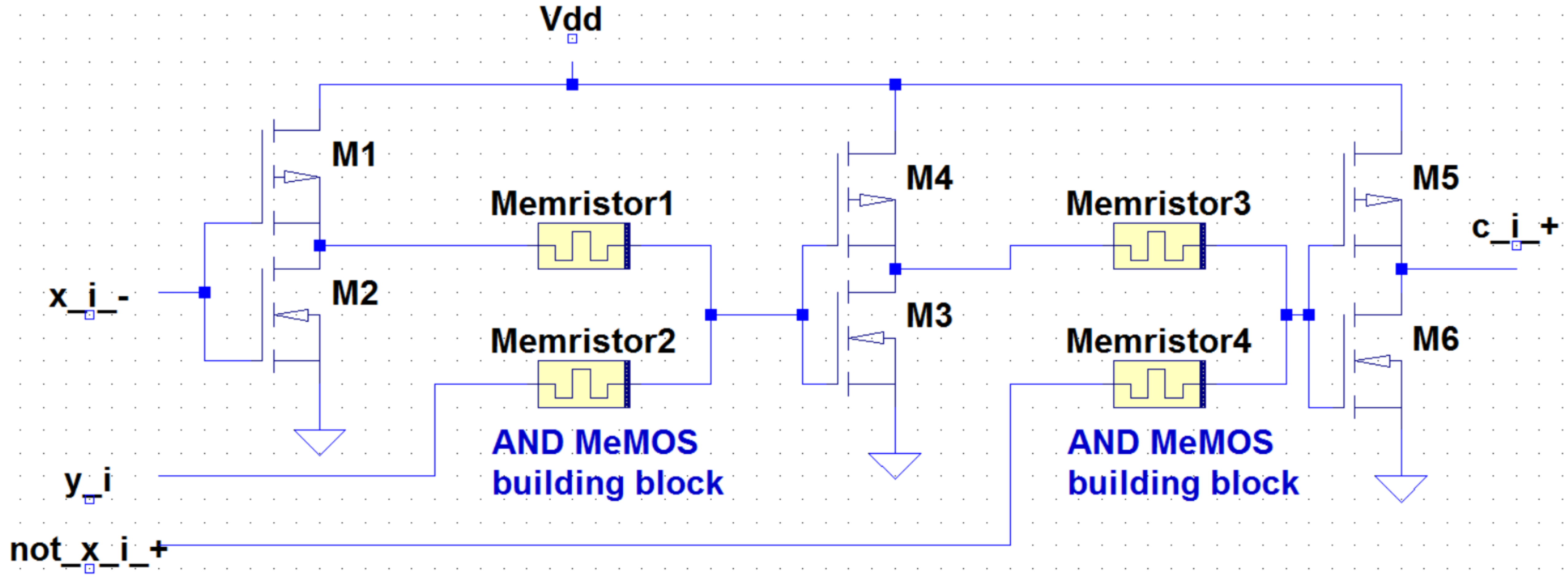}
\caption{Gate layout of MeMOS circuit for calculating $c_i^+$ signal (\ref{eq:first_c}). Shown are two AND building blocks in MeMOS. An OR block would be reversely connected to the inputs on left side.}
\label{fig:MeMOS_c_plus}
\end{figure}

The gained expression can be directly mapped in a 1-to-1 fashion onto a functionally equivalent MeMOS circuit shown in Fig. \ref{fig:MeMOS_c_plus}. We repeated this procedure for all other equations for both ternary adders, cascaded the gained circuits to build complete ternary adders as MeMOS circuits. These circuits have been simulated by an own written simulator.  

\section{Simulation system for MeMOS circuits}

In order to identify the possible processing speed and the necessary energy consumption of our found MeMOS circuits we are using simulation. For that we wrote a corresponding C++ program. The timely memristor behaviour was modelled by an implementation of the Euler solution of the differential equation to determine the memristance of the memristor devices. The same procedure is used in the TEAM and the VTEAM model published by Kvatinsky et.al. \cite{Kvatinsky_TEAM}, \cite{Kvatinsky_VTEAM}. In addition our simulation system accesses to the physical parameters of the memristors' given in the TEAM and VTEAM model.

Whereas the TEAM and the VTEAM model are deterministic models our simulation system considers also a statistical model that was introduced by Knowm \cite{Knowm}. The last model is of interest in the view of the fact that real, commercially available memristor devices are standing behind this model.

All these three models can be conveniently considered by our simulation system thanks to the object-orientated approach of C++. A memristor in our simulation system is a class, in which different functional behaviour can be instantiated by selecting a specific model. In order to simulate the functional behaviour of cascaded memristor circuits, as e.g. in the hybrid MeMOS logic, a tree of cascaded and interconnected memristors is built up. This tree is parsed and by that the memristor's inputs can be determined and the output is calculated in discrete time steps according to the Euler procedure. 

The CMOS inverters in MeMOS circuits necessary for signal restoring are functionally modelled as follows. If the input applied to such an inverter is more than 0.7 of the supply voltage, $Vdd$, then the complete voltage hub is switched through and the complete voltage supply level is accessible at the inverter's output. An analogue behaviour is assumed for the case that the input is below $0.3 \times V_{dd}$. In this case the output is pulled down to ground signal. By that it is possible by means with the parameters of the memristor models to evaluate MeMOS and in principle also other memristor circuits to determine latency and energy consumption by simulation.

Table \ref{table:Programm_extract} gives an excerpt of the listing of the source code for the simulation of a MeMOS circuit. It shows the definition of a MeMOS circuit and how it is possible to simulate this circuit with different models, just by defining one of the parameters TEAM, VTEAM, or KNOWM during the instantiation of a memristor MeMOS circuit (line 4). The MeMOS circuit itself is defined by the call of corresponding AND or OR memristor functions (line 3) which simulate the building blocks of MeMOS (line 1 and 2). The Boolean parameter at the end of the function call (line 3) determines if the output of the MeMOS circuit has to be inverted or not. 

\begin{table}[b]
\centering%
\begin{tabular}{cl}
       &  namespace MeMOS $\lbrace$ \\
       &  ... \\
     1 &  // defines MeMOS AND and OR building block  \\ 
       &  // AND has reversely interconnected poles to OR \\  
       &  typedef Gate<Mem\_pos, Mem\_neg> AND;\\
       &  typedef Gate<Mem\_neg, Mem\_pos> OR;\\
       &  \\
     2 &  // template for positive and negative memristor poles \\
       &  template<typename Mem\_pos, typename Mem\_neg>\\
       &  \\
     3 &  // Definition of MeMOS circuit for $c_i^+$ logic tree \\
       &  return create\_node(Input(x\_plus), create\_node\\
       &  (Inverter(y), Input(x\_minus), OR(par2, par1), true), \\
       &  OR(par2, par1), false); \\    
       &  ... \\
%       &  ... \\
     4 &  // defines MeMOS circuit for ternary adder with \\
       &  // TEAM model; if VTEAM or KNOWM shall be used \\
       &  // replace TEAM by VTEAM or KNOWM \\
       &  MeMOS::Radix\_2\_Step\_4\_Adder<Mem\_pos\_TEAM, \\
       &  Mem\_neg\_TEAM> adder(width, par1\_team, par2\_team);
\end{tabular}
\caption{Extract of the C++ simulation program. Syntactic definition of the circuit shown in Fig. \ref{fig:MeMOS_c_plus}}
\label{table:Programm_extract}
\end{table} 

With that program and the known values of a ternary CMOS adder attached to memristor registers, which have been publiseh in \cite{MemSysPaper}, we can now carry out a comparison study between ternary adders using MeMOS and conventional CMOS.

\section{Comparative evaluation of memristors} 

First we investigated the memristor behaviour after an initial reset of the memristor, i.e. the both memristors in the building blocks had a state variable 0.5. This corresponds to the situation that the memristance is exactly adjusted at the zero point of its I/V hysteresis curve. We tried to find out at which operating frequency a correct switching would fail because the ions have not enough time to move to another location in the memristor cavity to change the memristance clearly either to the high or to the low resistance mode. Table \ref{table:Perf_Eval_Reset_Memr} shows the gained results for an applied moderate voltage level of 1.8\,V for three different memristor models. 

Apparently is the much lower processing rate for the memristors described by the Knowm model. In this context is to say that Knowm memristors are the only commericially memristor whereas the other two models represent research memristors. Furthermore, the Knowm memristors are primarily thought for detection and learning applications based on neuromorphic processing schemes which can tolerate lower processing rates. It becomes clear that the other two models are more appropriate for implementing high-speed arithmetic. It is also interesting to see that the adder with three steps shows better latency values than the adder with four steps due to the higher possible operating frequency. That holds in particular very strongly for memristors described with the TEAM model. Even though the complexity of the gates in the $base\_2\_step\_4$ adder is lower than in the $base\_2\_step\_3$ adder the lower number of gates that have to be passed gives the advantage for the $base\_2\_step\_3$ adder. The values for the energy consumption refer to the processing of input operands with a digit width of 40. We will need these values later.

\begin{table}[b]
\begin{center}
\begin{tabular}{lll|lll}
\multicolumn{3}{c|}{Base\_2\_step\_3} & \multicolumn{3}{c}{Base\_2\_step\_4} \\ [0.5ex] % inserts table
%heading
\hline
\multirow{2}{9mm}{Team} & \multirow{2}{9mm}{Vteam} & \multirow{2}{9mm}{Knowm} & \multirow{2}{9mm}{Team} & \multirow{2}{9mm}{Vteam} & \multirow{2}{9mm}{Knowm} \\
\multirow{2}{9mm}{1.2ns} & \multirow{2}{9mm}{1.5ns} & \multirow{2}{9mm}{300$\mu$s} & \multirow{2}{9mm}{40ns} & \multirow{2}{9mm}{4ns} & \multirow{2}{9mm}{400$\mu$s} \\
\multirow{2}{9mm}{2.5GHz} & \multirow{2}{9mm}{2GHz} & \multirow{2}{9mm}{10KHz} & \multirow{2}{9mm}{100MHz} & \multirow{2}{9mm}{1GHz} & \multirow{2}{9mm}{160KHz} \\
\multirow{2}{9mm}{44pJ} & \multirow{2}{9mm}{885pJ} & \multirow{2}{9mm}{65$\mu$J} & \multirow{2}{9mm}{77pJ} & \multirow{2}{9mm}{2.25nJ} & \multirow{2}{9mm}{55$\mu$J} \\
 & & & & & \\
\hline
\end{tabular}
\caption{Performance evaluation of ternary adders after reset of memristors (static operation).}
\label{table:Perf_Eval_Reset_Memr}
\end{center}
\end{table}

However, in practice the behaviour would be worse. It is unlikely that after a switching process the same memristor's state will adjust to the same state  after a reset. In order to mimic a realistic scenario of cascaded MeMOS circuits which have to process different operand pairs one after the other, we determined randomly 1000 pairs of ternary operands that are given in as continuous data stream, each for a width of 8, 16, 32, and 64 digits. With the simulation system we calculated the average energy consumption per addition. As mentioned, the simulation software allows to specify an operation frequency. The reciprocal of this operating frequency corresponds to a certain amount of time that is applied to the memristor's inputs. 

It turned out by the simulation results that for all three models the energy consumption doubled if the digit width is doubled, too. This was to expect since the ternary adders have a complete regular setup concerning their logic blocks and the the spreading of a carry signal is clearly limited. The latency depends on the number of MeMOS blocks that have to be passed for the two ternary adders. We assume the reciprocal of the operating frequency, $f$, as run time through one MeMOS building block including therein contained CMOS inverters. Therefore, we need a latency of $3 \times 1/f$ for the $base\_2\_step\_3$ adder, and of $4 \times 1/f$ for the $base\_2\_step\_4$ adder.  

Table \ref{table:Perf_Eval_Dynamic_Memr} shows a kind of cut-off frequency, e.g. 350 MHz for the $base\_2\_step\_3$ adder, up to which we have got error-free simulation results. For higher frequencies the time the input signals were applied to the memristor inputs is not sufficient long to switch from a previous stored HIGH resistance to a LOW resistance or vice versa. It is interesting to see that the $base\_2\_step\_4$ adder shows a more stable behaviour than the $base\_2\_step\_3$ adder in the realistic dynamic situation. For the static case shown in Table \ref{table:Perf_Eval_Reset_Memr} this was reversed. Furthermore, it is important to mention that this high speed processing times could only be achieved if the voltage is increased to 6\,V for the TEAM and to 4\,V for the VTEAM model. This could possibly become a problem for a hybrid integration with CMOS since such high values are no more standard. With the simulated energy consumption values it is now also possible to determine the energy delay product for the ternary adders and to compare them with the CMOS solution that uses memristors as multi-value registers what is done in the next chapter. 

\section{Summary and conclusion}  

We investigated in this paper the realization of ternary adders using the possibilities of the MeMOS approach, i.e. combining logic circuits based on memristors with CMOS inverters used for signal refreshing. The presented ternary adders exploit the potential of multi-value memristors. 
 
Concerning the values for energy and latency the break even point for a ternary adder using multi-value memristor registers versus a CLA adder was about 40 digits according to the results in \cite{MemSysPaper}. At 40 digits the presented ternary adder $base\_2\_step\_3$ started to produce a better ED product, that was about 30 ns $ \times $ pJ. We looked in this paper if better results can be achieved with ternary MeMOS adders. We can fix that is clearly not the case for a dynamic switching mode with MeMOS. The CMOS ternary adder with memristor registers in \cite{MemSysPaper} showed a latency of 3\,ns and an energy consumption of 10\,pJ which is less than the corresponding values in Table \ref{table:Perf_Eval_Dynamic_Memr} for both MeMOS ternary adders. Concerning the static case the latency for the $base\_2\_step\_3$ offers with 1.2\,ns an improvement of 60\%. The energy delay product is with $1.2\,ns \times 44\,pJ = 52,8 ns \times pJ$ about 75\% worse. The ratio concerning energy is even more worse because the simulation does not consider so far the energy consumption of the inverters in the MeMOS circuits. 

Therefore we come to the following conclusion. For ternary adders it seems to be  better to prefer CMOS circuits with multi-value memristor registers at the moment. However, we will investigate in the future potentials to optimize the MeMOS circuits. E.g., it could be possible to combine multiple steps of the adders, which are now strictly separated. Furthermore, it is to investigate if the approach using MAD and MAGIC gates produces better results since they need less inverters.   
 
\begin{table}[!htb]
\begin{center}
\begin{tabular}{cc|cc}
\multicolumn{2}{c|}{Base\_2\_step\_3} & \multicolumn{2}{c}{Base\_2\_step\_4} \\ % [0.5ex] % inserts table
%heading
\hline
\multirow{2}{16mm}{TEAM} & \multirow{2}{16mm}{VTEAM} & \multirow{2}{16mm}{TEAM} & \multirow{2}{16mm}{VTEAM}  \\
% & & & \\
\multirow{2}{16mm}{350 MHz} & \multirow{2}{16mm}{100 MHz} & \multirow{2}{16mm}{500 MHz} & \multirow{2}{16mm}{100 MHz} \\
\multirow{2}{16mm}{8.5ns @6V} & \multirow{2}{16mm}{30ns @4V} & \multirow{2}{16mm}{8ns @6V} & \multirow{2}{16mm}{40ns @4V} \\
\multirow{2}{16mm}{38.8 pJ} & \multirow{2}{16mm}{240 nJ} & \multirow{2}{16mm}{22.5 pJ} & \multirow{2}{16mm}{295 nJ} \\
& & & \\
\hline
\end{tabular}
\caption{Performance evaluation of ternary adders processing subsequent data operands (dynamic operation).}
\label{table:Perf_Eval_Dynamic_Memr}
\end{center}
\end{table}
\section{Acknowledgement}
The author would like to thank his students Jonas Schmitt and Jonathan Martschinke who wrote the C++ simulation program. 

\bstctlcite{bstctl:etal, bstctl:nodash, bstctl:simpurl}
\bibliographystyle{IEEEtranS}

\newpage

\bibliography{references}

\end{document}